\documentclass[conference]{IEEEtran}
\IEEEoverridecommandlockouts
\usepackage{setspace}
\usepackage{cite}
\usepackage{amsmath,amssymb,amsfonts,mathtools,bm}
\usepackage{amsthm}
\usepackage{algorithm}
\usepackage{graphicx}
\usepackage{textcomp}
\usepackage{xcolor,float}
\usepackage{tabularx}
\usepackage{textcomp}
\usepackage{diagbox}
\usepackage{svg}
\usepackage{booktabs}
\usepackage{subfigure}
\usepackage{hyperref}

\usepackage{algpseudocode}
\usepackage{graphicx}
\usepackage{makecell}
\usepackage{amsthm}
\usepackage{sidecap}
\usepackage{microtype}
\usepackage{booktabs} 
\usepackage{multirow}
\usepackage{float}
\usepackage{adjustbox} 
\usepackage{colortbl}
\usepackage{makecell}
\usepackage{float}
\usepackage{url}
\usepackage{balance}
\usepackage{bbding}
\usepackage{hyperref}
\usepackage{microtype}
\usepackage{graphicx}
\usepackage{booktabs} 
\usepackage{multirow}
\usepackage{array}
\usepackage{makecell}

\usepackage{tikz}

\usepackage{changes}
\usepackage{ifthen}
\newboolean{showchanges}
\setboolean{showchanges}{false}  

\newcommand{\add}[1]{%
    \ifthenelse{\boolean{showchanges}}%
        {\textcolor{blue}{#1}}
        {#1\relax}
}

\definecolor{lime}{HTML}{A6CE39}
\DeclareRobustCommand{\orcidicon}{%
    \begin{tikzpicture}
    \draw[lime, fill=lime] (0,0) 
    circle [radius=0.16] 
    node[white] {{\fontfamily{qag}\selectfont \tiny ID}};    \draw[white, fill=white] (-0.0625,0.095) 
    circle [radius=0.007];    \end{tikzpicture}
    \hspace{-2mm}}
\foreach \x in {A, ..., Z}{%
    \expandafter\xdef\csname orcid\x\endcsname{\noexpand\href{https://orcid.org/\csname orcidauthor\x\endcsname}{\noexpand\orcidicon}}
    }

\allowdisplaybreaks
\renewcommand{\baselinestretch}{1}

\usepackage{orcidlink}  

\begin{document}

\title{Amortized Neural SVD for XL-MIMO: Structure-Guided Factor Prediction for Beamforming and Multi-Stream Utility}
\author{\IEEEauthorblockN{Yue Zhang\IEEEauthorrefmark{1},
Yiyan Zhang\IEEEauthorrefmark{2},
Ruijin Sun\IEEEauthorrefmark{1},
Honggang Jia\IEEEauthorrefmark{1},
Chen Gong\IEEEauthorrefmark{1}}
\IEEEauthorblockA{
\IEEEauthorrefmark{1}State Key Laboratory of ISN and School of Telecommunications Engineering, Xidian University, Xi'an, 710071, China\\
\IEEEauthorrefmark{2}School of Computer Science and Technology, Xi'an Jiaotong University, Xi'an, 710049, China\\
Email: \{25011210995, jiahg,25011211225\}@stu.xidian.edu.cn,zhangyy414@stu.xjtu.edu.cn, sunruijin@xidian.edu.cn }
}
    \maketitle

\IEEEdisplaynontitleabstractindextext

\IEEEpeerreviewmaketitle

\begin{abstract}
Singular value decomposition (SVD) is a core operation in multiple-input multiple-output (MIMO) beamforming, but the cubic complexity of standard SVD routines can lead to a major latency bottleneck as array dimensions scale to extremely large sizes. This paper presents a fully learned neural operator that avoids explicit SVD computation by directly mapping channel matrices to truncated low-rank factors for precoder and combiner design. In contrast to iterative numerical solvers and algorithm-unrolled networks, the proposed structure-aware model, termed \emph{SVDNet}, produces these factors in a single forward pass at inference, shifting the per-instance decomposition cost to offline training. The model also includes lightweight constraints to enforce basic algebraic properties required by beamforming, such as semi-unitarity of the singular vectors and nonnegative singular values, without invoking matrix factorization kernels. Experiments on extremely large-scale MIMO channels with matrix dimensions up to $512\times512$ show that the proposed approach achieves spectral efficiency close to exact SVD-based beamforming in single-stream transmission and consistently improves multi-stream sum-rate over representative learned baselines, indicating good scalability for low-latency wireless processing. Code is available at \url{https://github.com/ZY021023/Neural-SVD-for-MIMO}.
\end{abstract}


\section{Introduction}

Singular value decomposition (SVD) is a fundamental numerical primitive in multiple-input multiple-output (MIMO) signal processing. Many canonical transmission strategies can be expressed through the dominant singular modes of the channel matrix: in the practically prevalent single-stream regime, the principal right singular vector determines the optimal beamforming direction, while in multi-stream operation, the leading singular subspace enables spatial multiplexing and power allocation~\cite{a56g,A10baselin1,a9SVD,a12AI-enable}. As antenna arrays evolve toward massive and extremely large-scale MIMO (XL-MIMO), these SVD-based operations are invoked repeatedly across users and time–frequency resources, making the computational efficiency of SVD a first-order concern for both real-time adaptation and large-scale system evaluation~\cite{a11jsac,a4MIMO,a13xuemin6g,a3MassiveMIMO}.

\begin{figure}[t]
  \centering
  \includegraphics[width=\linewidth]{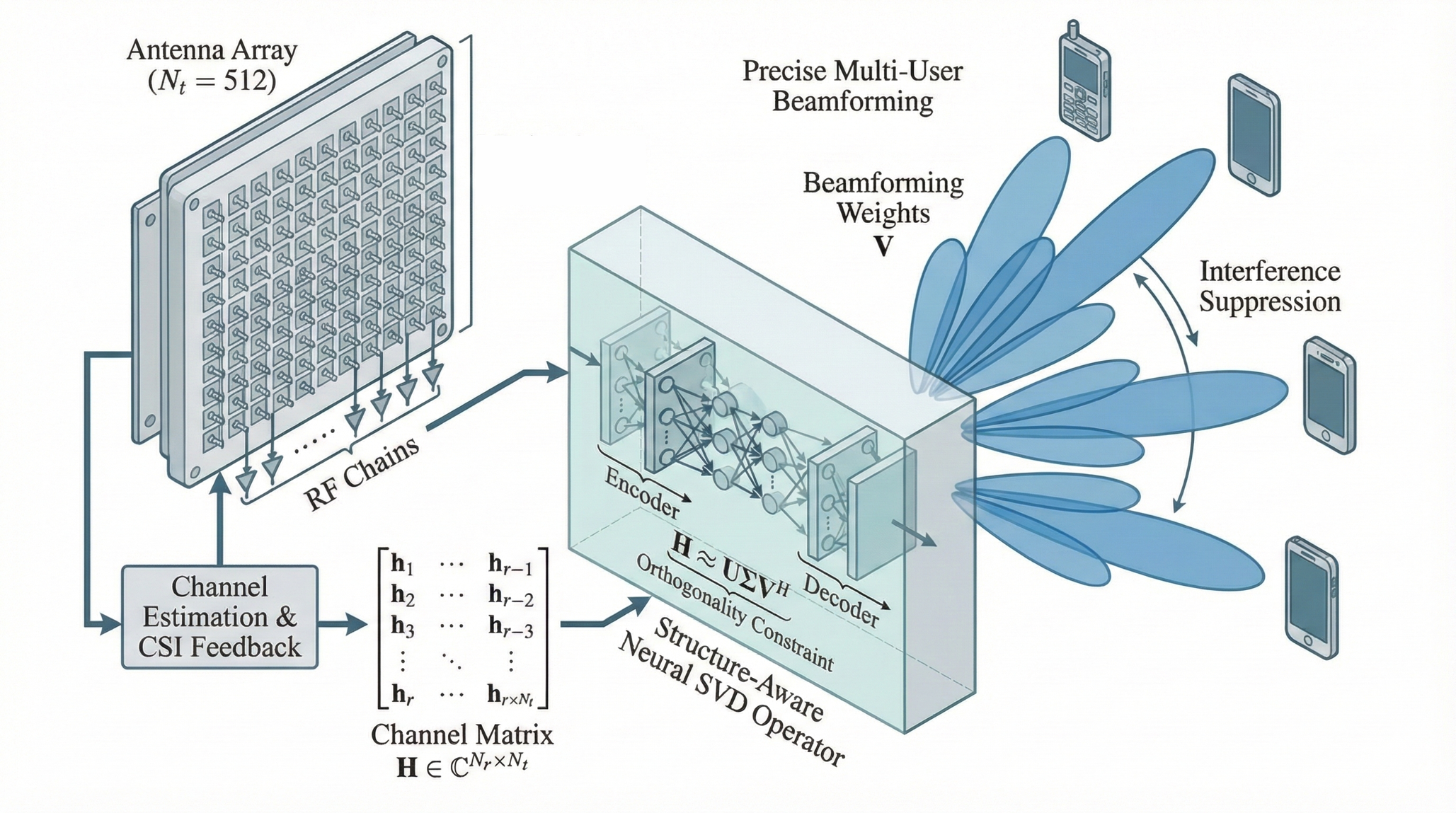}
  \caption{System overview of the proposed learned neural SVD operator.}
  \label{fig:scene_overview}
  \vspace{-0.6em}
\end{figure}

The need to accelerate SVD becomes acute as array dimensions scale. In standard MIMO pipelines, a general-purpose dense SVD is routinely invoked to extract dominant modes for each channel realization. However, for dense matrices, computing an exact SVD incurs steep arithmetic and memory costs: for an $N\times N$ matrix, the leading-order complexity is cubic, i.e., $\mathcal{O}(N^{3})$, accompanied by substantial $\mathcal{O}(N^{2})$ memory traffic that often becomes a major factor in wall-clock runtime as $N$ grows. Consequently, repeatedly calling an exact library routine (e.g., via LAPACK or \texttt{torch.svd}) can become a dominant latency bottleneck in both real-time processing loops and large-scale simulation frameworks~\cite{a8pytorch,a14dang2020should}.

While partial, iterative, or randomized methods can mitigate costs when only the top-$k$ subspace is required, their runtime depends on data-dependent convergence behavior. Moreover, these methods still involve repeated matrix-vector products and orthogonalization steps, which complicates predictable per-sample execution at extremely large dimensions. These considerations motivate an amortized, structure-preserving learned operator with fixed and controllable inference cost, evaluated rigorously against exact SVD baselines in terms of both communication utility and wall-clock scaling.

From a learning perspective, accelerating SVD is intrinsically challenging because the output is highly structured and utility-critical. Unlike generic matrix regression, valid SVD factors must satisfy stringent algebraic constraints-most notably, the unitarity of singular vectors and the nonnegativity and ordering of singular values-to remain actionable for downstream signal processing. Small violations of orthogonality can induce subspace leakage and unstable beamforming directions, translating into tangible rate loss even when the reconstruction error appears modest. This challenge is compounded in complex-valued channels, where phase consistency and unitary structure are essential. These considerations suggest that a viable learned substitute should be viewed not as an unstructured predictor of $\mathbf{H}$, but as a structured operator whose outputs remain valid under communication metrics~\cite{A10baselin1}.

Motivated by this viewpoint, we adopt an operator-learning approach: rather than solving SVD from scratch for each realization, we learn an amortized SVD operator that enforces structural validity by design. Specifically, as illustrated in Fig.~\ref{fig:scene_overview}, we propose a structure-aware neural SVD model, termed SVDNet, that maps a complex-valued channel matrix directly to structured factors $(\hat{\mathbf{U}}, \hat{\boldsymbol{\Sigma}}, \hat{\mathbf{V}})$ in a single forward pass. These factors can then be used by standard beamforming and precoding rules to form transmit/receive weights and evaluate downstream communication utility. The architecture is tailored to factorization, combining matrix-structured encoding with a lightweight, fixed-depth differentiable unitary-guidance mechanism to promote near-unitary singular-vector outputs, together with SVD-specific constraints such as nonnegativity and ordering consistency~\cite{a6attention,a15VAE}. We evaluate this learned operator using communication-centric utility metrics and wall-clock scaling behavior relative to an exact SVD reference. In the dominant single-stream regime, the network-only output achieves near-reference beamforming utility while providing substantial latency reductions, with the runtime advantage growing significantly as antenna dimensions scale up to $512\times512$. For multi-stream operation, we additionally introduce an optional predicted-subspace refinement as an accuracy--complexity knob, enabling improved utility without reverting to full SVD in the main loop. The main contributions of this paper are summarized as follows:
\begin{enumerate}
  \item Utility-aware learned SVD operator. We cast SVD in MIMO processing as a structured numerical operator and propose a learned model that predicts $(\hat{\mathbf{U}}, \hat{\boldsymbol{\Sigma}}, \hat{\mathbf{V}})$ for complex channel matrices in a single forward pass. The method is evaluated primarily through communication utility rather than solely reconstruction error.
  
  \item Structure-aware factorization design. We develop an SVD-specific architecture targeting structural validity. This includes matrix-structured encoding and a fixed-depth differentiable unitary-guidance component to stabilize singular vectors, together with constraints that enforce nonnegativity and consistent alignment of singular values and vectors.
  
  \item Scaling evidence in XL-MIMO-scale settings. Using an exact SVD routine as a reference, we provide systematic evaluation on $128\times128$, $256\times256$, and $512\times512$ channels under the current measurement-anchored setting. Results demonstrate near-reference single-stream beamforming utility and favorable wall-clock scaling trends with increasing speedups at larger dimensions, supported by an optional refinement extension for multi-stream operation.
\end{enumerate}

\section{System Model and Metrics}

\subsection{System Model and Reference Decomposition}
We consider a narrowband MIMO channel matrix $\mathbf{H} \in \mathbb{C}^{M \times N}$, where $M$ and $N$ denote the numbers of receive and transmit antennas, respectively. The singular value decomposition (SVD) of $\mathbf{H}$ is defined as
\begin{equation}
\mathbf{H} = \mathbf{U} \bm{\Sigma} \mathbf{V}^{\mathsf{H}},
\end{equation}
where $\mathbf{U} \in \mathbb{C}^{M \times r}$ and $\mathbf{V} \in \mathbb{C}^{N \times r}$ have orthonormal columns, $\bm{\Sigma} = \mathrm{diag}(\mathbf{s}) \in \mathbb{R}_+^{r \times r}$ with singular values $s_1 \ge s_2 \ge \cdots \ge s_r \ge 0$, and $r = \min(M, N)$. Throughout this paper, we use \texttt{torch.svd} as an exact reference decomposition.

Our goal is to replace per-instance numerical SVD with a learned amortized factorization operator. Given $\mathbf{H}$, the proposed neural model produces a low-rank factorization $(\hat{\mathbf{U}}, \hat{\mathbf{s}}, \hat{\mathbf{V}})$, where $\hat{\mathbf{U}} \in \mathbb{C}^{M \times R}$, $\hat{\mathbf{V}} \in \mathbb{C}^{N \times R}$, $\hat{\mathbf{s}} \in \mathbb{R}_+^{R}$, and $R \ll r$. The predicted factors are then directly used to construct beamforming/precoding directions and to evaluate communication utility against the reference SVD~\cite{a7SVDbeamfoming}.

\begin{figure*}[t]
  \centering
  \includegraphics[width=\textwidth]{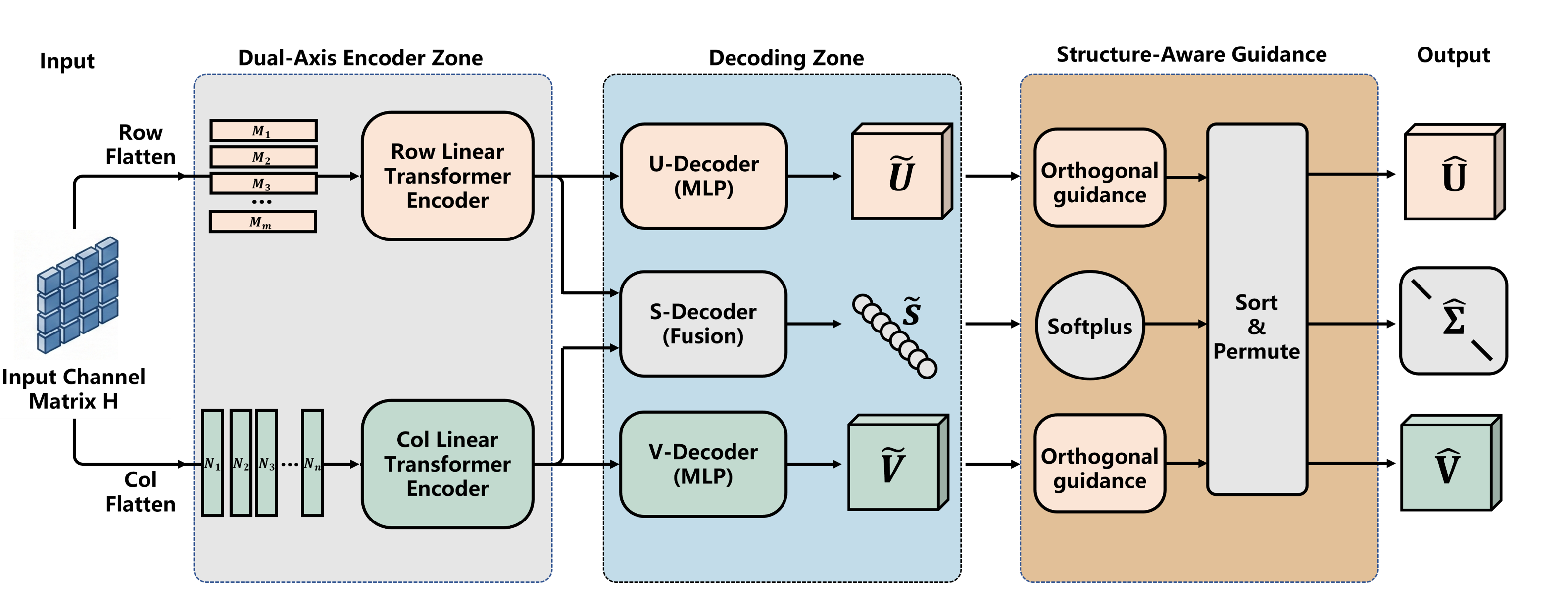}
  \caption{Architecture of the proposed learned neural SVD operator with structure-aware post-processing.}
  \label{fig:svdnet_arch}
  \vspace{-0.8em}
\end{figure*}

\subsection{Communication Utility and Scaling Latency Metrics}

\textbf{Single-stream (primary) utility:} We focus on the practically dominant single-stream beamforming setting ($k=1$), where performance is primarily determined by the principal right singular direction. Let $\mathbf{v} \in \mathbb{C}^{N}$ be a candidate transmit beamforming vector and $\bar{\mathbf{v}} = \mathbf{v} / \|\mathbf{v}\|_2$ its unit-norm normalization to ensure a consistent transmit-power convention. For a given linear SNR $\rho$, the rank-1 beamforming rate used in our evaluation is
\begin{equation}
R_1(\mathbf{H}, \bar{\mathbf{v}}) = \log_2 \left( 1 + \rho \|\mathbf{H} \bar{\mathbf{v}} \|_2^2 \right),
\end{equation}
where $\|\mathbf{H} \bar{\mathbf{v}} \|_2^2 = (\mathbf{H} \bar{\mathbf{v}})^{\mathsf{H}} (\mathbf{H} \bar{\mathbf{v}})$. We report the relative utility by comparing $R_1(\mathbf{H}, \hat{\mathbf{v}}_1)$ against the reference $R_1(\mathbf{H}, \mathbf{v}_1)$, where $\mathbf{v}_1$ is the principal right singular vector from \texttt{torch.svd}. We primarily present the normalized ratio $R_1(\mathbf{H}, \hat{\mathbf{v}}_1) / R_1(\mathbf{H}, \mathbf{v}_1)$ to isolate the performance loss induced by amortized inference.

\textbf{Multi-stream (secondary) utility:} For $k > 1$, we evaluate an interference-aware sum-rate induced by a set of $k$ transmit/receive directions. Denote $\mathbf{U}_k = \mathbf{U}(:, 1:k)$, $\mathbf{V}_k = \mathbf{V}(:, 1:k)$, and define the effective coupling matrix
\begin{equation}
\mathbf{A} = \mathbf{U}_k^{\mathsf{H}} \mathbf{H} \mathbf{V}_k \in \mathbb{C}^{k \times k}.
\end{equation}
If the factors were perfectly unitary and aligned with the true SVD, $\mathbf{A}$ would be diagonal. In practice, learned factors may yield off-diagonal terms representing residual inter-stream coupling. We treat diagonal entries $A_{ii}$ as intended gains and off-diagonal entries $A_{ij} (j \neq i)$ as interference leakage. With a power allocation vector $\mathbf{p} \in \mathbb{R}_+^k$ ($\sum p_i = 1$), the per-stream signal-to-interference-plus-noise ratio (SINR) is computed as
\begin{equation}
\mathrm{SINR}_i = \frac{\rho p_i |A_{ii}|^2}{\nu_i + \rho \sum_{j \neq i} p_j |A_{ij}|^2},
\end{equation}
where $\nu_i$ is the $i$-th diagonal entry of $\mathbf{U}_k^{\mathsf{H}} \mathbf{U}_k$ to account for non-unitary factors. The multi-stream sum-rate is $\sum_{i=1}^k \log_2(1 + \mathrm{SINR}_i)$.


\textbf{Latency and scaling:} To quantify solver-substitution benefits, we measure decomposition latency for (i) the learned amortized factorization and (ii) the exact reference SVD via \texttt{torch.svd}. We report the speedup ratio $t_{\mathrm{svd}} / t_{\mathrm{net}}$ and emphasize its scaling with dimensions $(M, N)$. The key motivation is that the amortized inference cost is bounded by fixed architectural choices (e.g., fixed output rank $R$ and fixed-depth structural guidance), leading to increasing advantages as array sizes grow.

\section{Proposed Method}
\subsection{Learned Neural SVD Operator}
We develop a learned neural SVD operator that amortizes per-instance decomposition into a single forward pass. As illustrated in Fig.~\ref{fig:svdnet_arch}, given a complex MIMO channel matrix $\mathbf{H}\in\mathbb{C}^{M\times N}$, the operator $\mathcal{F}_\theta$ predicts a low-rank SVD-like factorization $(\hat{\mathbf{U}},\hat{\mathbf{s}},\hat{\mathbf{V}})$, where $\hat{\mathbf{U}}\in\mathbb{C}^{M\times R}$, $\hat{\mathbf{V}}\in\mathbb{C}^{N\times R}$, $\hat{\mathbf{s}}\in\mathbb{R}_+^{R}$, and $R\ll \min(M,N)$. The predicted factors are consumed directly by standard beamforming/precoding rules (e.g., $\hat{\mathbf{v}}_1$ for $k=1$ and $\hat{\mathbf{V}}_{1:k}$ for $k>1$), enabling solver substitution in latency-critical pipelines.

The operator follows a compact three-stage design. A normalization stage first stabilizes channel magnitude variation using dataset-level statistics stored in the model. The encoder then extracts matrix-aware representations through row- and column-oriented branches aligned with the two natural axes of $\mathbf{H}$. Three factor heads decode these features into provisional low-rank factors $(\tilde{\mathbf{U}}, \tilde{\mathbf{s}}, \tilde{\mathbf{V}})$, which are subsequently converted by a lightweight structural post-processing stage into the final outputs $(\hat{\mathbf{U}}, \hat{\mathbf{s}}, \hat{\mathbf{V}})$. In particular, the vector factors are regularized by differentiable unitary guidance, while the singular values are constrained to be nonnegative and sorted in descending order. Deployment requires only a forward evaluation of $\mathcal{F}_\theta$; no explicit matrix factorization routines are used in the network-only inference path.

During training, we supervise the predicted factors through matrix reconstruction rather than direct singular-vector regression. Given a target channel matrix $\mathbf{H}_{\mathrm{gt}}$, we reconstruct
\[
\hat{\mathbf{H}}=\hat{\mathbf{U}}\,\mathrm{diag}(\hat{\mathbf{s}})\,\hat{\mathbf{V}}^{\mathsf H}.
\]
We define
\[
\mathcal{L}_{\mathrm{rec}}
=
\frac{\lVert \mathbf{H}_{\mathrm{gt}}-\hat{\mathbf{H}}\rVert_F}
{\lVert \mathbf{H}_{\mathrm{gt}}\rVert_F+\epsilon},
\]
\[
\mathcal{L}_{\mathrm{ortho}}
=
\sum_{\mathbf{Q}\in\{\hat{\mathbf{U}},\hat{\mathbf{V}}\}}
\lVert \mathbf{Q}^{\mathsf H}\mathbf{Q}-\mathbf{I}_R\rVert_F,
\]
and optimize
\begin{equation}
\mathcal{L}
=
\mathcal{L}_{\mathrm{rec}}
+
\lambda_{\mathrm{ortho}}\,\mathcal{L}_{\mathrm{ortho}}.
\end{equation}
During training, $\lambda_{\mathrm{ortho}}$ is gradually increased so that the early stage emphasizes low-rank reconstruction, while later epochs place stronger weight on semi-unitary structure. This curriculum-like weighting improves optimization stability by allowing the network to first capture the dominant matrix content and then refine the structural validity of the predicted factors. The above terms are averaged over the minibatch during training. This choice is also robust to the inherent phase ambiguity of singular vectors, since paired phase rotations of $\hat{\mathbf{U}}$ and $\hat{\mathbf{V}}$ leave the reconstructed matrix $\hat{\mathbf{H}}$ unchanged.
\subsection{Structure-Aware Design and Unitary Guidance}
Learning to output SVD factors is constrained by the fact that downstream performance depends on algebraic structure rather than reconstruction alone. In MIMO beamforming and precoding, singular vectors represent physically meaningful spatial directions; violations of unitarity may lead to unstable beamformers and, in multi-stream settings, residual inter-stream coupling that directly degrades achievable rate. Accordingly, the proposed operator injects SVD-specific inductive bias to produce factors that are both trainable and directly actionable.

The design is structure-aware in two complementary ways. First, it adopts a fixed low-rank parameterization with rank~$R$, matching the practical need for dominant modes while controlling inference cost. Second, it enforces SVD-compatible factor semantics through differentiable post-processing: predicted singular values are constrained to be nonnegative via a smooth positivity mapping, calibrated to the input energy scale (Frobenius-norm matching), and sorted in descending order with synchronized column permutation of $\hat{\mathbf{U}}$ and $\hat{\mathbf{V}}$. This ordering consistency is crucial in the dominant-mode regime to ensure that the first column corresponds to the strongest spatial direction across samples.

The remaining bottleneck is orthogonality: unconstrained decoders may produce vectors that are only approximately orthogonal and can drift from semi-unitarity at scale. We therefore incorporate a fixed-depth differentiable unitary guidance layer $(\hat{\mathbf{U}},\hat{\mathbf{V}})=\mathcal{P}(\tilde{\mathbf{U}},\tilde{\mathbf{V}})$, applied separately to the decoded factors. Let $X\in\mathbb{C}^{m\times R}$ denote the complex-valued output of the $U$- or $V$-decoder (with $m=M$ or $N$). We first normalize each column of $X$ and then perform $d$ iterations
\begin{equation}
  U_{t+1} \;=\; \tfrac{1}{2} U_t\bigl(3I_R - U_t^{H}U_t\bigr), 
  \quad t = 0,\ldots,d-1,
\end{equation}
where $U_0$ is the normalized input $X$, and $I_R$ denotes the $R\times R$ identity matrix. The output $\mathcal{P}(X)=U_d$ encourages $U_d^{H}U_d \approx I_R$ without invoking explicit matrix factorization routines. This module is used consistently during both training and inference.

\subsection{Multi-Stream Enhancement}
Although our primary emphasis is the dominant single-stream regime ($k=1$), the learned factors also support multi-stream transmission. For $k>1$, performance becomes increasingly sensitive to the quality of the predicted subspace and residual coupling among streams. We therefore employ an optional refinement restricted to a low-dimensional subspace predicted by the network. Given $\hat{\mathbf{V}}\in\mathbb{C}^{N\times R}$, we form $\mathbf{V}_{\text{sub}}=\hat{\mathbf{V}}_{[:,1:r_{\text{sub}}]}$ with $r_{\text{sub}}=\min\{\max(k,r_{\text{align}}),R\}$ (we use $r_{\text{align}}=3$ in the final experiments). A Rayleigh--Ritz update is performed in this subspace by constructing
\begin{equation}
\mathbf{G}=(\mathbf{H}\mathbf{V}_{\text{sub}})^{H}(\mathbf{H}\mathbf{V}_{\text{sub}})
=\mathbf{V}_{\text{sub}}^{H}\mathbf{H}^{H}\mathbf{H}\mathbf{V}_{\text{sub}},
\end{equation}
and extracting its $k$ dominant eigenvectors $\mathbf{E}_k$. The refined right-singular directions are $\mathbf{V}_{\text{ref}}=\mathbf{V}_{\text{sub}}\mathbf{E}_k$, followed by $\sigma_i=\|\mathbf{H}\mathbf{v}_i\|_2$ and $\mathbf{u}_i=\mathbf{H}\mathbf{v}_i/\sigma_i$ for $i=1,\ldots,k$. This step is not required for the dominant-mode case but improves stream decoupling for $k>1$ under the interference-aware sum-rate metric in Section~II.

\subsection{Computational Complexity}
For a square $n\times n$ channel, numerical SVD has cubic complexity $\mathcal{O}(n^3)$. In contrast, as illustrated in Fig.~\ref{fig:svdnet_arch}, the proposed encoder processes the matrix through separate row- and column-wise branches and applies linear attention without forming an explicit $n\times n$ attention matrix. Under fixed embedding width, the per-layer attention cost is linear in sequence length, so the end-to-end deployment cost is dominated by the matrix-shaped projections and decoders. With fixed embedding width, number of heads, encoder depth, output rank $R$, and guidance depth, the overall complexity therefore scales approximately as $\mathcal{O}(n^2)$. The optional refinement adds only a small overhead, $\mathcal{O}(n^2 r_{\text{sub}}+r_{\text{sub}}^3)$, with small $r_{\text{sub}}$.

\section{Experiments}

\subsection{Datasets, Experimental Setup, and Protocol}
We evaluate the proposed learned SVD operator on complex-valued square MIMO channel matrices with $M=N\in\{128,256,512\}$, and additionally include $N\in\{32,64\}$ for scalability benchmarking. The dataset is constructed via a two-stage procedure anchored by a measured seed set of approximately 30,000 realizations at $128\times128$ provided by Huawei Xi'an Research Institute. To enable consistent evaluation across dimensions, we synthesize datasets at other sizes by matching statistics estimated from the seed set, thereby preserving key second-order, low-rank, and large-scale structural characteristics across dimensions. Each sample consists of a clean ground-truth channel $\mathbf{Y}\in\mathbb{C}^{M\times N}$ and a noisy observation $\mathbf{X}=\mathbf{Y}+\mathbf{W}$, where $\mathbf{W}$ denotes additive circularly symmetric complex Gaussian noise with variance chosen to meet the target signal-to-noise ratio (SNR). This measurement-anchored construction is intended to evaluate scaling behavior in XL-MIMO-scale settings under the statistics induced by the measured seed set.

\begin{table}[t]
\centering
\caption{Normalized Spectral-Efficiency Ratio Under Water-Filling for Single- and Multi-Stream Cases}
\label{tab:rate_ratio}
\footnotesize
\setlength{\tabcolsep}{2.5pt}
\renewcommand{\arraystretch}{1.2}
\begin{tabular}{cccc}
\toprule
\multicolumn{4}{c}{Single-Stream ($k=1$): Pure Neural Output }\\
\midrule
$N$ & $\eta_1$ @ 0 dB & $\eta_1$ @ 10 dB & $\eta_1$ @ 20 dB \\
\midrule
128 & 0.9950 & 0.9965 & 0.9975 \\
256 & 0.9933 & 0.9952 & 0.9965 \\
512 & \textbf{0.9959} & \textbf{0.9969} & \textbf{0.9975} \\
\midrule
\multicolumn{4}{c}{Multi-Stream ($k=3$): \textsc{NN} / \textsc{NN+REF}}\\
\midrule
$N$ & $\eta_3$ @ 0 dB & $\eta_3$ @ 10 dB & $\eta_3$ @ 20 dB \\
\midrule
128 & 0.8611 / \textbf{0.9434} & 0.7923 / \textbf{0.9308} & 0.7138 / \textbf{0.9247} \\
256 & 0.8189 / \textbf{0.9227} & 0.7506 / \textbf{0.9149} & 0.6761 / \textbf{0.9154} \\
512 & 0.6954 / \textbf{0.8535} & 0.6145 / \textbf{0.8494} & 0.5359 / \textbf{0.8585} \\
\bottomrule
\end{tabular}
\end{table}

\begin{table}[t]
\centering
\caption{Inference latency and peak GPU memory allocation.}
\label{tab:lat_mem}
\scriptsize
\setlength{\tabcolsep}{2.2pt}
\renewcommand{\arraystretch}{1.15}
\begin{tabular}{c c c c c c c c c}
\toprule
& \multicolumn{3}{c}{CPU} & \multicolumn{3}{c}{GPU} & \multicolumn{2}{c}{Peak GPU Mem} \\
\cmidrule(lr){2-4}\cmidrule(lr){5-7}\cmidrule(lr){8-9}
$N$ & \textbf{SVDNet} & Exact SVD & Speedup & \textbf{SVDNet} & Exact SVD & Speedup & \textbf{SVDNet} & Exact SVD \\
& (ms) & (ms) & & (ms) & (ms) & & (MB) & (MB) \\
\midrule
128 & \textbf{6.09} & 19.02 & 3.12$\times$ & \textbf{4.99} & 9.11 & 1.83$\times$ & \textbf{48.6} & 48.6 \\
256 & \textbf{16.18} & 67.20 & 4.15$\times$ & \textbf{6.02} & 21.65 & 3.59$\times$ & \textbf{89.2} & 89.7 \\
512 & \textbf{27.88} & 156.39 & 5.61$\times$ & \textbf{4.67} & 77.58 & 16.60$\times$ & \textbf{175.0} & 178.0 \\
\bottomrule
\end{tabular}
\end{table}

\begin{table*}[t]
\centering
\caption{Comparison with learned decomposition baselines in terms of normalized spectral-efficiency ratio $\eta_k$.}
\label{tab:acc_baselines_k1k3}
\renewcommand{\arraystretch}{1.2}
\setlength{\tabcolsep}{8pt}
\small
\begin{tabular}{c l ccc ccc}
\toprule
\multirow{2}{*}{$N$} & \multirow{2}{*}{Method} &
\multicolumn{3}{c}{Single-stream ($k=1$)} &
\multicolumn{3}{c}{Multi-stream ($k=3$)} \\
\cmidrule(lr){3-5}\cmidrule(lr){6-8}
& & $\eta_1$ @ 0 dB & $\eta_1$ @ 10 dB & $\eta_1$ @ 20 dB &
$\eta_3$ @ 0 dB & $\eta_3$ @ 10 dB & $\eta_3$ @ 20 dB \\
\midrule

\multirow{3}{*}{32}
& \textbf{SVDNet} & \textbf{0.9948} & \textbf{0.9963} & \textbf{0.9976} & \textbf{0.9189} & \textbf{0.8608} & \textbf{0.7859} \\
& Peken~\cite{A10baselin1} & 0.9924 & 0.9948 & 0.9966 & 0.8254 & 0.7524 & 0.6678 \\
& AED~\cite{a2baselin2} & 0.8096 & 0.8396 & 0.8782 & 0.6106 & 0.4814 & 0.3505 \\
\midrule

\multirow{3}{*}{64}
& \textbf{SVDNet} & \textbf{0.9985} & \textbf{0.9989} & \textbf{0.9992} & \textbf{0.9605} & \textbf{0.9183} & \textbf{0.8570} \\
& Peken & 0.9953 & 0.9968 & 0.9977 & 0.7940 & 0.7073 & 0.6307 \\
& AED & 0.8268 & 0.8617 & 0.8994 & 0.5685 & 0.4262 & 0.3156 \\
\midrule

\multirow{3}{*}{128}
& \textbf{SVDNet} & \textbf{0.9950} & \textbf{0.9965} & \textbf{0.9975} & \textbf{0.8611} & \textbf{0.7923} & \textbf{0.7138} \\
& Peken & 0.9906 & 0.9938 & 0.9960 & 0.7199 & 0.6404 & 0.5461 \\
& AED & 0.8309 & 0.8632 & 0.8975 & 0.4803 & 0.3589 & 0.2809 \\
\midrule

\multirow{3}{*}{256}
& \textbf{SVDNet} & \textbf{0.9933} & \textbf{0.9952} & \textbf{0.9965} & \textbf{0.8189} & \textbf{0.7506} & \textbf{0.6761} \\
& Peken$^{\dagger}$ (DNC) & 0.0212 & 0.0306 & 0.0482 & 0.0042 & 0.0065 & 0.0103 \\
& AED & 0.8339 & 0.8652 & 0.8946 & 0.3131 & 0.2374 & 0.1293 \\
\midrule

\multirow{3}{*}{512}
& \textbf{SVDNet} & \textbf{0.9959} & \textbf{0.9969} & \textbf{0.9975} & \textbf{0.6954} & \textbf{0.6145} & \textbf{0.5359} \\
& Peken$^{\dagger}$ (DNC) & 0.0158 & 0.0253 & 0.0394 & 0.0036 & 0.0048 & 0.0067 \\
& AED & 0.8661 & 0.8918 & 0.9127 & 0.2747 & 0.2167 & 0.1785 \\
\bottomrule
\end{tabular}

\vspace{0.3em}
\footnotesize
\textit{Note:} $^\dagger$ indicates runs that converged to near-degenerate low-utility solutions under the common training protocol; these cases are marked as \textsc{DNC}.
\end{table*}

\noindent\textbf{Evaluation protocol.}
Unless otherwise stated, all learned methods infer factorization factors from the noisy observation $\mathbf{X}$, while communication utilities are evaluated on the corresponding clean channel $\mathbf{Y}$. As an upper bound, we compute the SVD of $\mathbf{Y}$ using \texttt{torch.svd}, which serves as the oracle reference for the normalized utility metrics reported below.

\noindent\textbf{Implementation details.}
For each dimension, we train a separate model on 20,000 samples (90\%/10\% train/validation split) and evaluate on a disjoint test set of 1,000 samples. Complex channel matrices are represented in concatenated real-valued form. During training, the noisy observation $\mathbf{X}$ is used as the network input, while the objective in Section~III is computed against the corresponding clean target channel $\mathbf{Y}$. The orthogonality weight $\lambda_{\mathrm{ortho}}$ is linearly increased during training, from $0.01$ at the beginning to $2.0$ at the final epoch. We evaluate SNRs of $\{0,10,20\}$\,dB and consider both single- and multi-stream transmission with $k\in\{1,2,3,4\}$ under water-filling power allocation; the main reported results focus on $k=1$ and $k=3$. The network outputs a rank-$R$ factorization with $R=32$. For multi-stream operation, we enable the optional subspace refinement (Section~III-C) within a predicted right-singular subspace of dimension $r_{\text{align}}=3$. We also examined the sensitivity of the unitary-guidance depth over $d\in\{1,2,3,4\}$; shallower settings ($d=1,2$) reduced downstream utility, whereas $d=4$ brought no consistent gain and slowed training convergence. All reported results therefore use $d=3$.

\noindent\textbf{Metrics and latency.}
Communication performance is quantified by the interference-aware achievable sum-rate defined in Section~II, computed from the effective channel $\mathbf{A}=\mathbf{U}_k^{\mathsf{H}}\mathbf{Y}\mathbf{V}_k$. For computational efficiency, we report decomposition-only latency for the learned operator and for \texttt{torch.svd} under identical measurement settings on an NVIDIA RTX 4060 GPU. The reported latency focuses on the core neural decomposition, while the optional refinement involves only low-dimensional operations.

\subsection{Experiment Results}
We report results for the single-stream regime ($k=1$) and the multi-stream regime ($k=3$), using water-filling power allocation based on predicted singular values. All methods are evaluated under the interference-aware achievable sum-rate in Section~II. We report the normalized spectral-efficiency ratio $\eta_k$ relative to the oracle SVD on the clean channel $\mathbf{Y}$. Table~\ref{tab:rate_ratio} reports $\eta_k$ for both regimes; the multi-stream block is shown as NN-only / NN+REF, where NN+REF denotes the optional refinement within the predicted subspace.

\vspace{2pt}
\noindent\textbf{Single-stream results.}
Table~\ref{tab:rate_ratio} (top block) shows that SVDNet achieves near-oracle dominant-mode utility across all tested dimensions and SNRs. In particular, $\eta_1$ remains within $[0.9933, 0.9975]$ for $N\in\{128,256,512\}$ over $\{0,10,20\}$\,dB, indicating that the learned operator provides a reliable surrogate for dominant-stream precoding.

\vspace{2pt}
\noindent\textbf{Multi-stream results and optional refinement.}
Table~\ref{tab:rate_ratio} (bottom block) shows that multi-stream transmission is more sensitive to structural errors that induce residual inter-stream interference. The raw neural output (NN-only) degrades with both $N$ and SNR; for example, at $N=512$, $\eta_3$ drops from $0.6954$ at 0\,dB to $0.5359$ at 20\,dB. The optional NN+REF stage substantially reduces this gap by re-solving the top-$k$ modes within the predicted $r_{\text{align}}$-dimensional subspace ($r_{\text{align}}=3$), improving $\eta_3$ from $0.5359$ to $0.8585$ at $N=512$ and 20\,dB. This indicates that the network captures a useful low-dimensional signal subspace, while the refinement provides a practical accuracy--latency trade-off without reverting to full SVD.

\vspace{2pt}
\noindent\textbf{Efficiency and scaling.}
Table~\ref{tab:lat_mem} compares SVDNet with the exact SVD baseline (\texttt{torch.svd}) in terms of decomposition-only latency on CPU/GPU and peak GPU memory allocation. Across both backends, the speedup increases with $N$. On GPU, SVDNet latency remains within $4.67$--$6.02$\,ms, whereas \texttt{torch.svd} increases to $77.58$\,ms at $N=512$, corresponding to a $16.60\times$ speedup. On CPU, the speedup reaches $5.61\times$ at $N=512$. Peak memory remains comparable across methods, indicating that the acceleration comes from computational efficiency rather than increased memory usage. This per-sample metric is also conservative: under batched inference (e.g., 8 samples), the effective GPU throughput speedup reaches approximately $140\times$.

\subsection{Accuracy Comparison with Learned Decomposition Baselines}
To benchmark scalability and architectural efficacy, we compare SVDNet with two representative learned decomposition baselines: \textit{Peken}, a pure neural operator with the same end-to-end formulation, and \textit{AED}, a prior-aided method based on QR decomposition~\cite{a2baselin2,A10baselin1}. All methods are trained and evaluated under the same protocol across $N\in\{32,\dots,512\}$. Table~\ref{tab:acc_baselines_k1k3} reports the normalized spectral-efficiency ratio $\eta_k$ relative to the oracle SVD, where larger values indicate better communication utility. To isolate the quality of the learned operators, the multi-stream block ($k=3$) reports raw neural outputs without optional refinement.

\begin{itemize}
    \item \textbf{Stability under dimension scaling (vs. Peken~\cite{A10baselin1}):} 
    Both SVDNet and Peken aim to learn the SVD operator directly from channel data. However, as shown in Table~\ref{tab:acc_baselines_k1k3}, Peken exhibits severe performance degradation at $N \ge 256$, where $\eta_1$ drops to very low values. This suggests that defining the problem correctly is not sufficient by itself; without explicit structural constraints, generic networks struggle to maintain stable optimization in high-dimensional factor spaces. In contrast, SVDNet incorporates unitary guidance to regularize the search space, enabling stable convergence and near-oracle performance even at extreme dimensions ($N=512$).
    
    \item \textbf{Comparison with the prior-aided baseline (vs. AED~\cite{a2baselin2}):} 
    While AED utilizes a valid mathematical prior (QR decomposition) to improve stability, it consistently underperforms SVDNet, especially in the multi-stream regime ($k=3$). This suggests that its underlying neural architecture---despite the QR backend---is less effective at capturing complex MIMO channel features. The rigidity of the QR block, combined with a generic feature extractor, limits its expressivity. By contrast, SVDNet employs a specialized, structure-aware architecture that explicitly models the interactions between singular vectors and values, leading to higher utility than the prior-aided baseline.
\end{itemize}

Overall, Table~\ref{tab:acc_baselines_k1k3} shows that SVDNet offers better optimization stability under dimension scaling than the pure neural baseline, while also achieving higher utility than the prior-aided baseline. Across $N=32$--$512$, it delivers near-oracle single-stream accuracy and consistently stronger multi-stream performance.

\section{Conclusion}
We proposed a structure-aware learned SVD operator for large-scale MIMO channels that amortizes matrix decomposition into a single forward pass. The proposed design yields near-oracle dominant-mode accuracy across a wide range of array sizes and remains effective in the multi-stream setting with an optional subspace refinement. Comparisons with learned decomposition baselines show that the gains do not arise from neuralization alone: a pure neural baseline collapses at large dimensions, while a prior-aided amortized baseline remains consistently less accurate. These results demonstrate that incorporating SVD structure into the network is key to scalable and stable learned decomposition for massive MIMO.

\bibliography{ref}
\bibliographystyle{IEEEtran}
\ifCLASSOPTIONcaptionsoff
  \newpage
\fi
\end{document}